\documentclass{ws-procs975x65}

\usepackage{epstopdf}

\allowdisplaybreaks

\begin{document}

\title{Baryonic Matter in the Hidden Local Symmetry Induced from Holographic QCD Models\footnote{Contribution to SCGT12 "KMI-GCOE Workshop on Strong Coupling Gauge
 Theories in the LHC Perspective", 4-7 Dec. 2012, Nagoya University.}}

\author{
Yong-Liang Ma$^\dag$, 
Ghil-Seok Yang$^\ddagger$, 
Yongseok Oh$^\ddagger$, 
Masayasu Harada$^\dag$, 
Hyun Kyu Lee$^{\dagger\dagger}$, 
Byung-Yoon Park$^\ast$ 
and Mannque Rho$^{\dagger\dagger,\S}$}

\address{$^\dag$Department of Physics, Nagoya University,  Nagoya,
464-8602, Japan \\
$^\ddagger$Department of Physics, Kyungpook National University,
Daegu 702-701, Korea \\
$^{\dagger\dagger}$Department of Physics, Hanyang University, Seoul 133-791, Korea \\
$^\ast$Department of Physics, Chungnam National University, Daejeon 305-764, Korea \\
$^\S$Institut de Physique Th\'eorique,
CEA Saclay, 91191 Gif-sur-Yvette c\'edex, France}

\begin{abstract}
Baryonic matter is studied in the Skyrme model by taking into account the roles of 
$\pi,$ $\rho$, and $\omega$ mesons through the hidden local symmetry up to 
$\mathcal{O}(p^4)$ terms including the homogeneous Wess-Zumino (hWZ) terms. 
Using the master formulas for the low energy constants derived from holographic 
QCD models the skyrmion matter properties can be quantitatively calculated with the input 
values of the pion decay constant $f_\pi$ and the vector meson mass $m_\rho^{}$. 
We find that the hWZ terms are responsible for the repulsive interactions of 
the $\omega$ meson. 
In addition, the self-consistently included $\mathcal{O}(p^4)$ terms with the 
hWZ terms is found to increase the half skyrmion phase transition point above 
the normal nucleon density.
\end{abstract}

\keywords{Baryonic matter; Skyrme model; Hidden local symmetry}

\bodymatter

\section{Introduction}

\label{sec:Intr}

Understanding the QCD phase structure is an extremely important problem in particle 
and nucleon physics, but is highly nontrivial because of the nonperturbative nature of 
strong interactions. Therefore, any theoretical investigation of the QCD phase structure requires effective
models for strong interactions at low energies. In this talk, we discuss the QCD phase structure for zero temperature baryonic matter in the Skyrme model based on the hidden local symmetry (HLS) up to $\mathcal{O}(p^4)$ 
terms including the homogeneous Wess-Zumino (hWZ) terms~\cite{Ma:2012kb,Ma:2012zm,Ma:2013}.
To uniformly describe the baryon, baryonic matter and also the change of hadron properties
in baryonic matter background, we apply the Skyrme's conjecture~\cite{Skyrme:1961vq} and
describe baryons as topological solitons in a mesonic theory. We also put the skyrmions into the face-centered-cubic (FCC) crystal~\cite{Lee:2003aq} for studying baryonic matter. 

Although the original Skyrme model is quite simple as it is just the nonlinear sigma model 
stabilized by the Skyrme term, it can reasonably describe the properties of a single nucleon
in vacuum in the sense of large $N_c$ expansion~\cite{Zahed:1986qz}. 
Since the original Skyrme model contains only pion degree of freedom, 
it is natural to improve the model predictions by including more dynamical degrees of 
freedom. When we increase the energy scale, the first resonance entering is the ground state $\rho$ and $\omega$
vector mesons. In the present work, we adopt the Hidden Local Symmetry (HLS) approach which is 
gauge equivalent to the nonlinear sigma model for including vector mesons~\cite{Harada:2003jx}.

Once extra degrees of freedom are added to an effective theory, the dynamics generally 
becomes more complicated by introducing more independent low energy constants (LECs) which
are to be determined by empirical data. However, it is not possible to determine all the LECs phenomenologically without ambiguity as the number of LECs increases, and this hinders the systematic investigation of the baryon and baryonic matter properties. In Ref.~\citenum{Ma:2012zm} we derived general master formulas for the LECs of HLS from a kind of holographic QCD (hQCD) models. This allows us to investigate the skyrmion and hadronic matter properties with only two input values of $f_\pi$ and $m_\rho$.

In Refs.~\citenum{Ma:2012kb,Ma:2012zm} we applied the HLS up to $\mathcal{O}(p^4)$ terms 
including the hWZ terms to investigate the single-skyrmion properties. 
It was found that the hWZ terms are responsible for the role of the $\omega$ meson 
and that the $\omega$ and $\rho$ mesons have very different roles indicated by the 
attractive nature of the $\rho$ meson and the repulsive interaction from the $\omega$ meson. 
As the next step for studying the role of vector mesons, we apply the above model to baryonic 
matter~\cite{Ma:2013} to study the skyrmion matter properties and the half skyrmion phase 
transition point.

This paper is organized as follows. In Sec.~\ref{sec:skyrmion} we investigate the single skyrmion properties from the HLS Lagrangian after discussing the implications of the infinite towers of vector mesons in the skyrmion properties. 
We then study the skyrmion matter properties from HLS in Sec.~\ref{sec:matter}, where we also compare our results with other model calculations. Section~\ref{sec:remark} contains a summary and perspective of our work.

\section{Skyrmion properties from the hidden local symmetry}

\label{sec:skyrmion}

We start with the study of the single skyrmion properties based on the HLS 
up to $\mathcal{O}(p^4)$ terms including the hWZ terms. 
Therefore, this model includes the $\pi$, $\rho$ and $\omega$ mesons.
The LECs of the HLS Lagrangian can be determined from the master formulas 
derived from hQCD models which allows us to control the LECs so that the skyrmion properties can be calculated in a self-consistent way.

The role of the infinite tower structure in five-dimension hQCD models in the skyrmion properties have 
been addressed in the large 't Hooft coupling $\lambda$ limit~\cite{Sutcliffe:2011ig}. 
It was found that, when only the pion was kept in the truncated model, the skyrmion 
has an excessive energy of about 25\% of the Bogomol'nyi bound, while the excess
energy is reduced to 7\% when the $\rho$ meson is included. 
This observation indicates that high-lying vector mesons make the theory 
flow to a conformal theory. 
However, in the large $\lambda$ limit, the theory does not include the repulsive 
interactions due to the $\omega$ meson which is the corrections of the next order in
$\lambda$, where the Chern-Simons (CS) term brings in a U(1) degree of freedom 
responsible for the $\omega$ meson and its tower.

To include the $\omega$ meson, we consider the hWZ terms in the HLS Lagrangian,
which allows us to study the role of the $\omega$ meson in the baryon properties. 
In addition, since the validity of the Skyrme model is supported by the large $N_c$ limit, 
for consistency, we need to take into account the higher order terms such as the $\mathcal{O}(p^4)$ 
terms in addition to the $\mathcal{O}(p^2)$ terms. 
Our Lagrangian is then summarized as
\begin{eqnarray}
\mathcal{L}_{\rm HLS} & = & \mathcal{L}_{\rm (2)} + \mathcal{L}_{\rm (4)} 
+ \mathcal{L}_{\rm anom} ,
\label{eq:Lag_HLS}
\end{eqnarray}
where $\mathcal{L}_{\rm (2)}, \mathcal{L}_{\rm (4)}$ and $\mathcal{L}_{\rm anom}$ 
stand for the $\mathcal{O}(p^2)$, $\mathcal{O}(p^4)$ and the hWZ terms, respectively. 
The explicit expressions for the Lagrangian  (\ref{eq:Lag_HLS}) responsible for the 
calculation of the skyrmion properties can be found in Ref.~\citenum{Ma:2012kb}.

The Lagrangian (\ref{eq:Lag_HLS}) has seventeen LECs, namely, $f_\pi$, $g$, $a$, 
$y_1 \sim y_9$, $z_4$, $z_5$ and $c_1 \sim c_3$. 
These LECs can be determined from the general master 
formula~\cite{Ma:2012kb,Ma:2012zm} derived from a class of hQCD models which 
is defined by the following general 5D action:
\begin{eqnarray}
S_{\rm 5} & = & S_{\rm 5}^{\rm DBI} + S_{\rm 5}^{\rm CS},
\label{eq:actionhQCD}
\end{eqnarray}
where the Dirac-Born-Infeld part $S_{\rm 5}^{\rm DBI}$ and the
CS part $S_{\rm 5}^{\rm CS}$ are expressed as
\begin{eqnarray}
S_{\rm 5}^{\rm DBI} & = & 
N_c G_{\rm YM}\int d^4 x dz
\bigg\{ - \frac{1}{2} K_1(z) \mbox{Tr} \left[ \mathcal{F}_{\mu\nu} 
\mathcal{F}^{\mu\nu} \right] 
+ K_2(z) M_{KK}^2 \mbox{Tr} \left[ \mathcal{F}_{\mu z} 
\mathcal{F}^{\mu z} \right] \biggr\},
\label{eq:DBI} \\
S_{\rm 5}^{\rm CS} & = & \frac{N_c}{24\pi^2} \int_{M^4\times R} w_5^{} (A),
\label{eq:SSCS}
\end{eqnarray}
where $G_{\rm YM} \equiv \lambda/(108\pi^3)$ and 
$\mathcal{F}_{MN} = \partial_M \mathcal{A}_N - \partial_N  \mathcal{A}_M
- i [\mathcal{A}_M,\mathcal{A}_N]$ is the field strength of the 5D gauge field $\mathcal{A}_M$.
Here, we use the index $M = (\mu,z)$ with the convention $\mu = 0,1,2,3$ and $K_{1,2}(z)$ 
are the metric functions constrained by the gauge/gravity duality. 
In Eq.~(\ref{eq:SSCS}), $M^4$ and $R$ stand for the four-dimensional Minkowski space-time
and $z$-coordinate space, respectively, and $w_5^{} (A)$ is the CS 5-form.
Note that the DBI part is of $\mathcal{O}(\lambda^1)$ while the CS term is of $\mathcal{O}(\lambda^0)$. 
From the action (\ref{eq:actionhQCD}), after integrating out all the modes except the zero 
and the lowest-lying vector modes, the LECs of HLS are expressed as
\begin{eqnarray}
f_{\pi}^2 & = & N_c G_{\rm YM}^{} M_{KK}^2  \int dz K_2(z) \left[ \dot{\psi}_0^{}(z)
\right]^2, \quad  f_{\pi}^2 = N_c G_{\rm YM}^{} M_{KK}^2  \lambda_1^{} \langle \psi^2_1 \rangle, \nonumber\\
\frac{1}{g^2} & = & N_c G_{\rm YM}^{} \langle \psi_1^2 \rangle , \quad
y_1^{} = -y_2^{} = -N_c G_{\rm YM}^{} \left\langle \left(1 + \psi_1 - \psi_0^2 \right)^2
\right\rangle , \nonumber\\
y_3^{} & = & -y_4^{} = -N_c G_{\rm YM}^{} \left\langle \psi^2_1 \left(1 + \psi_1^{} \right)^2
\right\rangle , \quad 
y_5^{} = 2 y_8^{} = -y_9^{} = -2N_c G_{\rm YM}^{} \left\langle \psi_1^2 \psi_0^2 \right\rangle ,
\nonumber\\
y_6^{} & = & - \left( y_5^{} + y_7^{} \right) ,\quad
y_7^{} = 2N_c G_{\rm YM}^{} \left\langle \psi_1^{} \left ( 1 + \psi_1^{} \right)
\left(1 + \psi_1^{} - \psi_0^2 \right) \right\rangle ,
\nonumber\\
z_4^{} & = & 2N_c G_{\rm YM}^{} \left\langle \psi_1^{} \left( 1+\psi_1^{} - \psi_0^2 \right)
\right\rangle , \quad 
z_5^{} = - 2N_c G_{\rm YM}^{} \left\langle \psi_1^2 \left( 1 + \psi_1^{} \right) \right\rangle ,
\nonumber\\
c_1^{} & = &  \left\langle\!\!\left\langle
\dot{\psi}_0^{} \psi_1^{} \left( \frac{1}{2} \psi_0^2 + \frac{1}{6} \psi_1^2
- \frac{1}{2} \right) \right\rangle\!\!\right\rangle ,
\nonumber\\
c_2^{} & = & \left\langle\!\!\left\langle
\dot{\psi}_0^{} \psi_1^{} \left( -\frac{1}{2} \psi_0^2 + \frac{1}{6} \psi_1^2
+ \frac{1}{2} \psi_1^{} + \frac{1}{2} \right) \right\rangle\!\!\right\rangle, \quad
c_3^{} = \left\langle\!\!\left\langle
\frac{1}{2}\dot{\psi}_0^{} \psi_1^{2} \right\rangle\!\!\right\rangle ,
\label{eq:lecshls}
\end{eqnarray}
where $\lambda_1$ is the smallest (non-zero) eigenvalue of the eigenvalue equation,
\begin{equation}
- K_1^{-1} (z) \partial_z \left[ K_2(z) \partial_z \psi_n^{}(z) \right] =
\lambda_n \psi_n^{}(z)  ,
\label{eq:eigenpsi}
\end{equation}
and
\begin{eqnarray}
\langle A \rangle & \equiv & \int_{-\infty}^{\infty}  dz K_1(z) A(z), 
\qquad 
\langle\langle A \rangle\rangle \equiv  \int_{-\infty}^\infty dz A(z).
\end{eqnarray}

With the help of the formulas in Eq.~(\ref{eq:lecshls}), sixteen of the seventeen 
parameters can be fixed from the two parameters $G_{\rm YM}$ and $M_{\rm KK}$
which, in turn, can be determined by the pion decay constant $f_\pi$ and the 
vector meson mass $m_\rho$. 
The only undetermined parameter is the HLS-parameter $a$ or equivalently the
coupling constant $g$ as they are connected by the mass relation 
$m_\rho^2 = a g^2f_\pi^2$. 
From the point view of the 5D holographic model, this ambiguity is attributed to 
the normalization of the 5D wave function $\psi_1(z)$ that cannot be determined 
from the \textit{homogeneous} eigenvalue equation (\ref{eq:eigenpsi}). 
In Refs.~\citenum{Ma:2012kb,Ma:2012zm} we have shown that the parameter $g$ 
disappears from the expressions for the skyrmion mass if we express the LECs 
in terms of the normalized $\psi_1(z)$, which leads to the independence of the
skyrmion properties on the value of the HLS parameter $a$. 
This conclusion was numerically verified in Ref.~\citenum{Ma:2012zm}.

For numerical calculations, we use the empirical values, $m_\rho = 775.5$~MeV and 
$f_\pi = 92.4$~MeV as input values. 
And for the hQCD model, we take the Sakai-Sugimoto model~\cite{Sakai:2004cn}, 
where $K_1(z) = (1+ z^2)^{-1/3}$ and $K_2(z) = 1+z^2$. 
Here, we consider three versions of the HLS model: HLS$(\pi,\rho,\omega)$, 
HLS$(\pi,\rho)$ and HLS$(\pi)$.
HLS$(\pi,\rho,\omega)$ is the full model including the pion, $\rho$, and $\omega$ 
mesons and HLS$(\pi,\rho)$ is the model without the hWZ terms.
When the vector fields are integrated out, we have the original Skyrme model but 
with the Skyrme parameter determined by the hQCD model.
This is called HLS$(\pi)$. 
Our numerical results are summarized in Table~\ref{table:numphy} that shows
the skyrmion mass $M_{\rm sol}$, the $\Delta$-$N$ mass splitting $\Delta_M$, 
the winding number rms radius $\sqrt{\langle r^2 \rangle_W}$ and the energy rms radius 
$\sqrt{\langle r^2 \rangle_E}$.
For a comparison, we also show the results of ``the minimal model'' which includes the 
$\omega$ meson in a minimal way~\cite{Meissner:1986js} in the column of
HLS$_{\rm min}(\pi,\rho,\omega)$.
\begin{table}
\tbl{\label{table:numphy}
Skyrmion properties calculated from the HLS models. 
The soliton mass $M_{\rm sol}$ and the $\Delta$-$N$ mass difference $\Delta_M$ are 
in units of MeV, while $\sqrt{\langle r^2 \rangle_W^{}}$ and $\sqrt{\langle r^2 \rangle_E^{}}$ 
are in units of fm.}
{\begin{tabular}{@{}ccccccccc@{}}
\hline
\hline
& HLS$(\pi,\rho,\omega)$ & HLS$(\pi,\rho)$  & HLS$(\pi)$  & HLS$_{\rm min}(\pi,\rho,\omega)$ \\ \hline
$M_{\rm sol} $ & 1184 & 834 &  922 &  1407 \\
$\Delta_M$ & 448 & 1707 & 1014 & 259 \\ \hline
$\sqrt{\langle r^2 \rangle_W^{}}$ &  0.433 & 0.247 &  0.309 &  0.540 \\
$\sqrt{\langle r^2 \rangle_E^{}}$ &  0.608 &  0.371 &  0.417 & 0.725 \\
\hline
\hline
\end{tabular}}
\end{table}

Close inspection of the results given in Table~\ref{table:numphy} leads to the 
following observations.
Namely, the results from HLS$(\pi)$ and HLS$(\pi,\rho)$ explicitly show the 
attractive interaction nature arising from the $\rho$ meson, while
comparison of the results of HLS$(\pi,\rho)$ and HLS$(\pi,\rho,\omega)$ explicitly 
illustrates the repulsive nature due to the $\omega$ meson. 
In addition, the differences between the results from HLS$(\pi,\rho, \omega)$ 
and HLS$_{\rm min}(\pi,\rho,\omega)$ come from the $\mathcal{O}(p^4)$ terms and the
extra terms in the hWZ terms.

\section{Skyrmion matter from the hidden local symmetry}

\label{sec:matter}

\begin{figure}[t] %
\begin{center}
\includegraphics[scale=0.45]{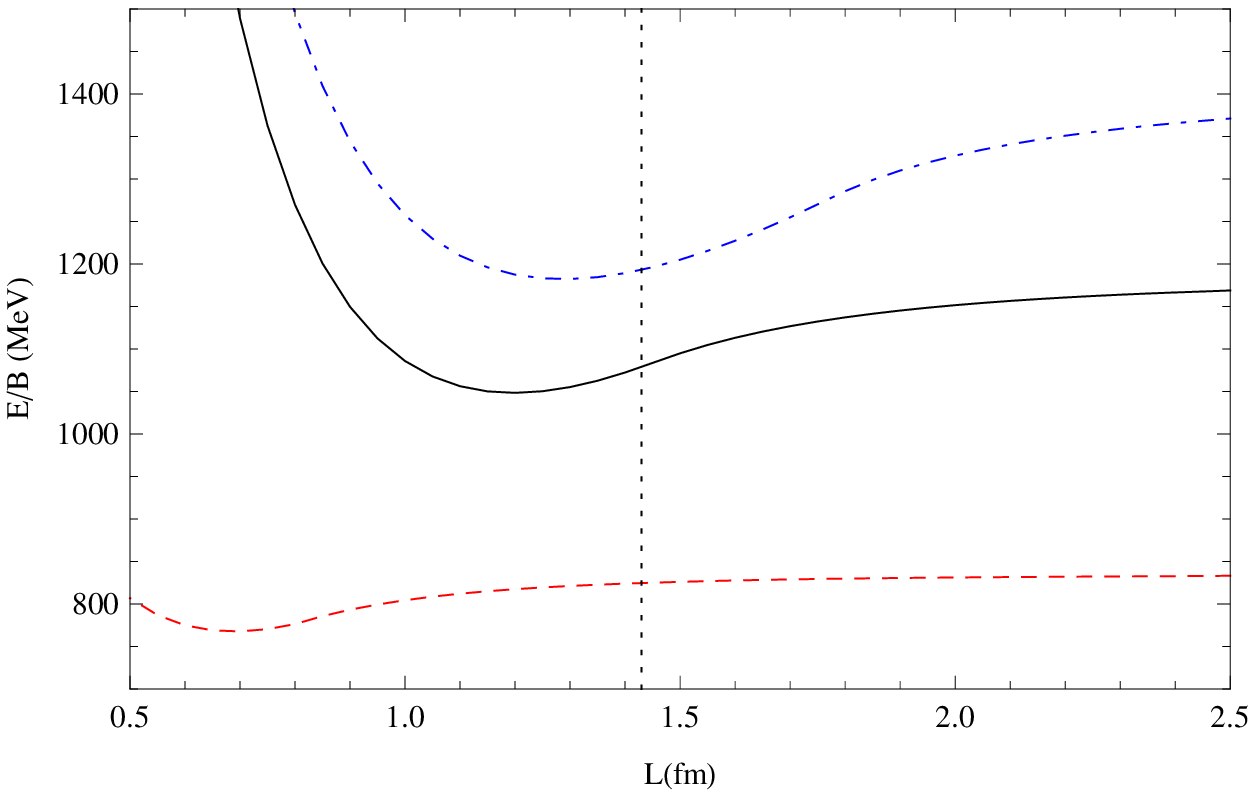} \quad 
\includegraphics[scale=0.45]{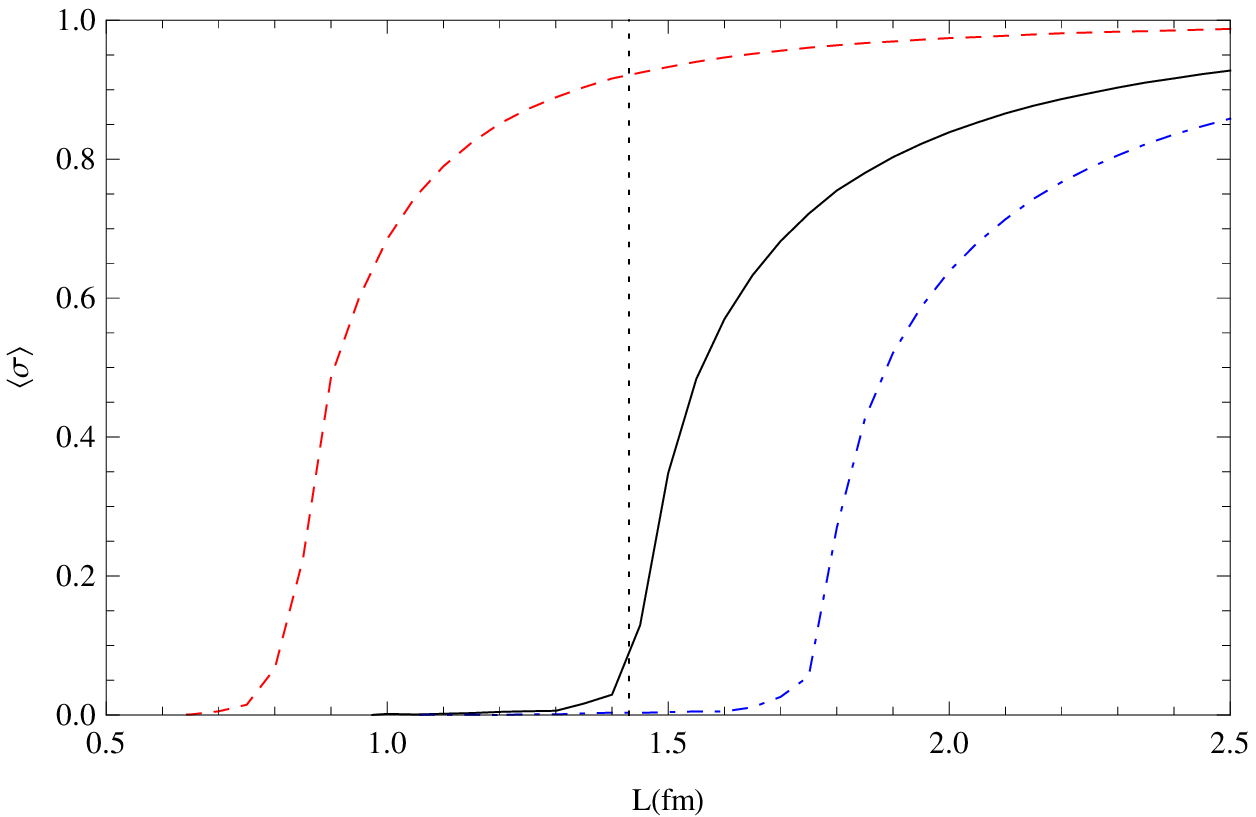}
\caption[]
{The energy per baryon number $E/B$ and the expectation
value $\langle \sigma\rangle$ of the minimum energy configuration 
for a given value of $L$. } 
\label{fig:compareSSmin} 
\end{center}
\end{figure}

In this Section, we apply the HLS model of the previous Section for the study of 
the skyrmion matter properties~\cite{Ma:2013} using the procedure 
explained in Ref.~\citenum{Lee:2003aq}, i.e., by putting skyrmions into the FCC crystal. 
Shown in Fig.~\ref{fig:compareSSmin} is the numerical results for $E/B$ and 
$\langle \sigma\rangle$ as functions of the FCC size parameter $L$. 
Here, $\langle\sigma \rangle$ is the average of the chiral field $U$ over the space 
where a single skyrmion occupies, which vanishes in the half-skyrmion phase. 
Here, we present the numerical results for  HLS$(\pi,\rho,\omega)$, HLS$(\pi,\rho)$ and 
HLS$_{\rm min}(\pi,\rho,\omega)$ by the solid, dahed, and dash-dotted lines, respectively. 
The position of the normal nuclear matter density is denoted by the vertical dotted line.

One can see that, in HLS$(\pi,\rho,\omega)$, the half-skyrmion phase transition 
happens almost at the normal nuclear matter density $n_0$. 
Comparing to that of HLS$_{\rm min}(\pi,\rho,\omega)$, we see that the inclusion 
of the $\mathcal{O}(p^4)$ interaction terms and the other $\pi$-$\rho$-$\omega$ 
interactions through the hWZ terms is crucial to make $n_{1/2}$ larger by 
the factor of $1.7$.
This is quite a noticeable improvement from the previous work~\cite{Park:2003sd},
where the position of the phase transition could not be explicitly pinned down. 
The higher value of $n_{1/2}$ of HLS$(\pi,\rho,\omega)$ compared to the 
HLS$_{\rm min}(\pi,\rho,\omega)$ result can be understood from the fact that 
the size of the single skyrmion is smaller in the former and the additional interactions 
in HLS$(\pi,\rho,\omega)$ weaken the repulsive interactions from the $\omega$ meson. 
This analysis is supported by the results from HLS$(\pi, \rho)$. 

In addition, we find the weak dependence of $E/B$ on the density in HLS$(\pi, \rho)$. 
This again emphasizes the role of the vector mesons. 
As we can see from Table~\ref{table:numphy} as well as from the conclusion of 
Ref.~\citenum{Sutcliffe:2011ig}, when only the pions 
are considered, the skyrmion carries a lot of excessive energy, while the inclusion of the 
$\rho$ meson reduces the soliton mass a lot and almost saturates the Bogomol'nyi bound. 
Therefore, only very little interaction energy remains in the skyrmion made of pion and the
$\rho$, and consequently it yields the (nearly) flat dashed curve for $E/B$.

\begin{figure}[t] %
\begin{center}
\includegraphics[scale=0.7]{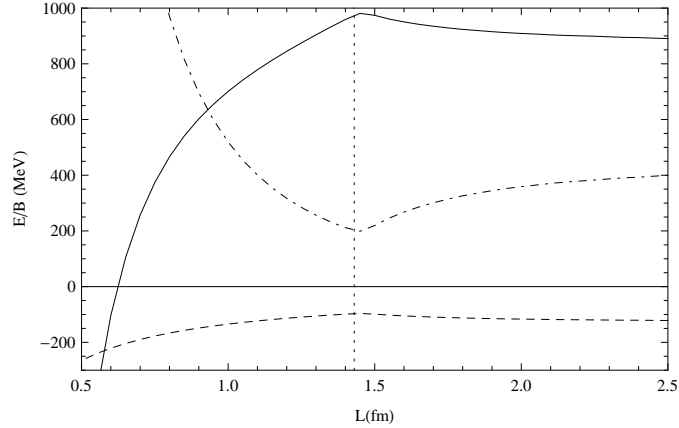}
\caption[]
{The energy per baryon number is plotted as a function of $L$. 
The solid, dashed and dash-dotted lines are $(E/B)_{\mathcal{O}(p^2)}$, 
$(E/B)_{\mathcal{O}(p^4)}$ and $(E/B)_{\rm hWZ}$, respectively.} 
\label{fig:compareterm} 
\end{center}
\end{figure}

In Fig.~\ref{fig:compareterm}, we present the skyrmion energy from each term 
in Eq.~(\ref{eq:Lag_HLS}) as a function of $L$. 
The plot shows that $(E/B)_{\mathcal{O}(p^4)}$ is negative for all the values of 
density, which explicitly shows that the inclusion of the higher order terms and 
vector mesons provide attractions and lead to the weak density dependence. 
On the contrary, in the single skyrmion phase, $(E/B)_{\mathcal{O}(p^2)}$ is dominant 
and the ratio $(E/B)_{\mathcal{O}(p^2)}/(E/B)_{\rm hWZ}$ even 
increases from $\sim 2 $ to $\sim 3$ as the density increases. 
However, after the half-skyrmion phase transition, $(E/B)_{\mathcal{O}(p^2)}$ starts to 
decrease, while $(E/B)_{\rm hWZ}$ starts to increase. 
Then, at some density the repulsive interaction due to the $\omega$ meson becomes 
the dominant interaction.

\section{Remarks and perspectives}

\label{sec:remark}

In the present work, we discuss the skyrmion and baryonic matter properties using the 
hidden local symmetry Lagrangian up to $\mathcal{O}(p^4)$ terms including 
the hWZ terms. 
Thanks to the holographic QCD models, the LECs can be determined by the
master formulas and the skyrmion properties can be calculated self-consistently
using two input parameters $m_\rho$ and $f_\pi$. 
This self-consistent approach was also recently applied to study the bound state 
picture of heavy baryons~\cite{Harada:2012dm}.

Due to the quantitative predictive power of the present framework, we apply this model 
to investigate the skyrmion matter properties by putting the skyrmions into the FCC crystal. 
For the crystal size dependence of the skyrmion energy, we found that, in 
HLS$_{\rm min}(\pi, \rho, \omega)$ and HLS$(\pi,\rho,\omega)$, the minimum of the 
skyrmion energy $n_{\rm min}$ appears at the same density and $n_{\rm min}$ 
in HLS$(\pi,\rho)$ appears at a very deep region. 
A noticeable result is that, in contrast to HLS$_{\rm min}(\pi, \rho, \omega)$, 
$n_{1/2}$ of HLS$(\pi,\rho,\omega)$ and HLS$(\pi,\rho)$ appears at density higher 
than the normal nuclear density $n_0$, which is consistent with our expectation.

It will be also interesting to study the role of the dilaton field in our HLS model and
the change of hadron properties in baryonic matter.
Work in this direction is in progress and will be reported elsewhere.

\section*{Acknowledgments}

The work reported here was partially supported by the WCU project of Korean
Ministry of Education, Science and Technology (R33-2008-000-10087-0).
The work of M.H. and Y.-L.M. was supported in part by Grant-in-Aid for
Scientific Research on Innovative Areas (No. 2104) ``Quest on New
Hadrons with Variety of Flavors'' from MEXT.
Y.-L.M. was supported in part by the National Science Foundation of 
China (NNSFC) under Grant No. 10905060.
The work of M.H. was supported in part by the Grant-in-Aid for
Nagoya University Global COE Program
``Quest for Fundamental Principles in the Universe: from Particles
to the Solar System and the Cosmos'' from MEXT, the JSPS
Grant-in-Aid for Scientific Research (S) $\sharp$ 22224003, (c)
$\sharp$ 24540266.
The work of Y.O. and G.-S.Y. was supported in part by the Basic Science Research
Program through the National Research Foundation of Korea (NRF) funded by
the Ministry of Education, Science and Technology (Grant \mbox{No.} 2010-0009381).

\bibliographystyle{ws-procs975x65}
\bibliography{ws-pro-sample}

\begin{thebibliography}{9}

\bibitem{Ma:2012kb} 
Y.-L.~Ma, Y.~Oh, G.-S.~Yang, M.~Harada, H.~K.~Lee, B.-Y.~Park and M.~Rho,
Phys.\ Rev.\ D {\bf 86}, 074025 (2012).  

\bibitem{Ma:2012zm} 
  Y.-L.~Ma, G.-S.~Yang, Y.~Oh, M.~Harada,
  Phys.\  Rev.\ D {\bf 87}, 034023 (2013).  

\bibitem{Ma:2013} 
Y.-L.~Ma {\it et al.\/},
to appear. 

\bibitem{Skyrme:1961vq} 
T.~H.~R.~Skyrme,
Proc.\ Roy.\ Soc.\ Lond.\ A {\bf 260}, 127 (1961).  

\bibitem{Lee:2003aq}  
H.-J.~Lee, B.-Y.~Park, D.-P.~Min, M.~Rho and V.~Vento,
Nucl.\ Phys.\ A {\bf 723}, 427 (2003).  

\bibitem{Zahed:1986qz} 
I.~Zahed and G.~E.~Brown,
Phys.\ Rept.\  {\bf 142}, 1 (1986).  

\bibitem{Harada:2003jx}
  M.~Harada and K.~Yamawaki,
  Phys.\ Rept.\  {\bf 381}, 1 (2003).

\bibitem{Sutcliffe:2011ig} 
  P.~Sutcliffe,
  JHEP {\bf 1104}, 045 (2011).  

\bibitem{Sakai:2004cn} 
T.~Sakai and S.~Sugimoto,
  Prog.\ Theor.\ Phys.\  {\bf 113}, 843 (2005).  

\bibitem{Park:2003sd} 
  B.-Y.~Park, M.~Rho and V.~Vento,
  Nucl.\ Phys.\ A {\bf 736}, 129 (2004).  

\bibitem{Meissner:1986js} 
  U.~G.~Meissner, N.~Kaiser and W.~Weise,
 Nucl.\ Phys.\ A {\bf 466}, 685 (1987).  
 
\bibitem{Harada:2012dm} 
  M.~Harada and Y.-L.~Ma,
Phys.\  Rev.\ D {\bf 87}, 056007 (2013).



\end{thebibliography}

\end{document}